\documentclass[reprint,aps,showkeys,nofootinbib]{revtex4-2}

\usepackage{graphicx}% Include figure files
\usepackage{dcolumn}% Align table columns on decimal point
\usepackage{bm}% bold math
\usepackage{hyperref}% add hypertext capabilities
\usepackage{amsmath}
\usepackage{amssymb}
\usepackage{amsfonts}
\usepackage{resizegather}
\usepackage{caption}
\usepackage[british]{babel}
\usepackage{physics}
\usepackage{esint} %For extended integrals (ointegral etc)

\usepackage{color}
\usepackage{ulem}
\usepackage{cancel}
\usepackage[export]{adjustbox}
\usepackage[newitem,newenum,defblank]{paralist}

\newcommand{\lmatt}{\mathcal{L}_\text{matt}}

\begin{document}

%\title{Finally: the matter Lagrangian of an ideal fluid is the energy density!}
% \title{The matter Lagrangian of an ideal fluid is its total energy density and not its pressure!}
\title[Matter Lagrangian of an ideal fluid]{The hidden truth of ideal
relativistic fluids: the matter Lagrangian is its total 
energy density, not its pressure!}

%\title{General Post-Newtonian Parametrisation in spherical symmetry at
%second order perturbation.}

\author{Sergio Mendoza}
\email{sergio@astro.unam.mx}
\affiliation{Instituto de Astronom\'{\i}a, Universidad Nacional Aut\'onoma de M\'exico, AP 70-264, Ciudad de M\'exico 04510, M\'exico}
\author{Sarah\'{\i} Silva}
\email{sgarcia@astro.unam.mx}
\affiliation{Instituto de Astronom\'{\i}a, Universidad Nacional Aut\'onoma de M\'exico, AP 70-264, Ciudad de M\'exico 04510, M\'exico}

%\date{Last updated 2020 ; in original form}
\date{\today}


\begin{abstract}
It has been established in the literature that the matter Lagrangian of
an ideal fluid can be expressed either as its total energy density or as
its pressure. In this work, we demonstrate that identifying the matter
Lagrangian with the pressure leads to physical inconsistencies, which are
resolved when the fluid is coupled to the  gravitational
field. In such a scenario, the matter Lagrangian necessarily assumes the
value of the total energy density. We thus conclude that, for an ideal
fluid, the only physically consistent choice for the matter Lagrangian
is its total energy density.
\end{abstract}

\keywords{General relativity and gravitation; Alternative gravity theories}

\maketitle

\section{Introduction}
\label{introduction}

  As mentioned by \citet{mendoza2021matter}, over the years it has become
a fact that the value of the matter Lagrangian for an ideal fluid given
by either  its total energy density, i.e. the sum of rest energy
density plus its internal energy density, or by its pressure 
yield the same energy momentum tensor in relativistic theories of
gravity.   Our intention in this article is to show that the pressure
value of the matter Lagrangian of an ideal fluid is incorrect and that
the only allowed value is the total internal energy of the fluid.

We define the matter action \( S \) as follows \citep[see e.g.][]{landau-fields}:

\begin{equation}
  S = \pm \int{ \lmatt \, \sqrt{-g} \, \mathrm{d}^4 x },
\label{action}
\end{equation}

\noindent where \( \lmatt \) represents the matter Lagrangian and \( g
\) is the determinant of the metric tensor \( g_{\alpha\beta} \).

  The Hilbert energy-momentum tensor is defined as~\citep[see
e.g.][]{landau-fields,harko-lobo-book,franklin17,nuastase19}:

\begin{equation}
  T_{\alpha\beta} =\pm \frac{ 2 }{ \sqrt{-g} }\frac{ 
    \delta \left( \sqrt{-g} \, \lmatt 
    \right) }{ \delta g^{\alpha\beta}},
\label{energy-momentum}
\end{equation}

\noindent or equivalently:

\begin{equation}
 T_{\alpha\beta} = \pm 2 \frac{ \delta \lmatt }{ \delta g^{\alpha\beta} } \mp
   g_{\alpha\beta} \lmatt.
\label{em-co}
\end{equation}

 Throughout the article we use Greek space-time indices from \( 0 \) to \(
4 \), Einstein's summation convention, a \((+,-,-,-)\) signature for the
metric tensor, units where the velocity of light
$c=1$ and as usual, we assume that the matter Lagrangian is independent
of the spatial derivatives of the metric tensor~\citep{landau-fields}.
The 4-velocity vector \( u_\alpha \) is normalised such that:

\begin{equation}
  u^\alpha u_\alpha = 1.
\label{normalu}
\end{equation}

  The article is organised as follows.  In section~\ref{hydro} we
describe the basic concepts of an ideal fluid used throughout the article
and explain how the value of the matter Lagrangian yields the total
internal energy density as explained by~\citet{mendoza2021matter}.  In
section~\ref{non-coupled-potential-flow} we describe how a potential
flow non-coupled with the gravitational field yields the pressure as the
value of the matter Lagrangian for an ideal fluid.
Section~\ref{coupled-potential-flow} shows that if a potential flow is
coupled to the gravitational field, the matter Lagrangian for an ideal
fluid is its total energy density.  Later on, in
Section~\ref{general-result} we show some very general hydrodynamical
results that show how the only possible solution to the matter
Lagrangian is its internal energy density.  Additionally it will be clear
in Section~\ref{discussion} that the well known result of the 
nullity of the variations of Taub's current, which immediately yields the
fact that the matter Lagrangian of an ideal fluid is its pressure, is
incorrect, showing additionally that the correct non-null variations of this
current yield the total energy density as the value of the matter
Lagrangian of a perfect fluid.  Also, in Section~\ref{discussion} we
discuss our results and show that an ideal fluid for which its matter
Lagrangian is given by the pressure cannot exist in nature.

\section{Hydrodynamics of an ideal fluid}
\label{hydro}

  An ideal fluid moves
  adiabatically~\citep{tooper,landau-fluids,rezzolla2013relativistic} so that the first law of
thermodynamics is given by:

\begin{equation}
    \mathrm{d} \left( \frac{ e }{n } \right) = - p \mathrm{d} \left(
      \frac{1}{n} \right),
    \label{flif}
\end{equation}

\noindent or equivalently:

\begin{equation}
  \mathrm{d}\left( \frac{ \omega }{ n } \right) = \frac{1}{n} \mathrm{d}p,
    \label{flw}
\end{equation}

\noindent where \( e, \  p, \  \) and \(  n  \) represent the total
(rest plus internal) energy density, the pressure and the particle number
density of the fluid.  The enthalpy per unit volume \( \omega \) is
given by $ \omega :=e+p$.  
In here and in what follows we use \citet{landau-fluids} formulation where extensive thermodynamic quantities
are measured per unit volume.

An ideal fluid satisfies a polytropic relation given by
\cite{tooper1965adiabatic, landau-fluids, landau2013statistical,
chandrasekhar1957introduction}:

\begin{equation}
    p=K n ^\gamma,
    \label{eof}
\end{equation}

\noindent where $K$ is a constant and so, the equation of state of an ideal fluid is given by~\cite{tooper1965adiabatic}:

\begin{equation}
    e = m n  + \frac{p}{\gamma - 1},
    \label{eof1}
\end{equation}

\noindent where $m$ is average mass per particle and $\gamma$ is a
constant equal to the ratio of the fluid's specific heats. The first term in equation~\eqref{eof1} 
is the rest energy density and the second one is its pure internal
energy density per unit volume which is obtained from equation~\eqref{eof}.

With the aid of equation~\eqref{eof1}, the first law of thermodynamics for an ideal fluid becomes:

\begin{equation}
        \frac{\mathrm{d} p}{ p} = \gamma \frac{\mathrm{d}n}{n}.
        \label{fleos}
\end{equation}

  The energy-momentum tensor \( T_{\alpha\beta} \) for an ideal fluid is
given by~\citep{landau-fluids}:

\begin{equation}
  T_{\alpha\beta} = \left( e+p \right) u_\alpha u_\beta - p
  g_{\alpha\beta},
\label{energy-momentum-ideal}
\end{equation}

\noindent and satisfies the conservation equation:

\begin{equation}
  \nabla_{\alpha} T^{\alpha\beta} = 0.
\label{conservation}
\end{equation}

The contraction of equation~\eqref{conservation} with the 4-velocity \(
u_\beta \) and with the use of equations~\eqref{energy-momentum-ideal}
and~\eqref{flif} yields the continuity equation:

\begin{equation}
  \nabla_\alpha \left( n u^\alpha \right)  = 0.
\label{continuity}
\end{equation}

Contraction of equation~\eqref{conservation} with the projection tensor 
$h_{\mu \beta}:= g_{\mu \beta}-u_{\mu}u_{\beta}$  yields Euler's
equation given by:

\begin{equation}
    \omega u^{\mu}\nabla_{\mu}u_{\nu}- h^{\mu}_{\nu}\nabla_{\mu}p=0,
\label{euler2}
\end{equation}

\noindent which can be rewritten as:

\begin{equation}
    u^\alpha \nabla_\alpha \left( \frac{\omega}{n} u_\beta \right) =
\nabla_{\beta} \left( \frac{\omega}{n}  \right)
\label{ec-t-c}
\end{equation}

\noindent with the help of equation~\eqref{flw}.  The previous relation
can be rewritten as:

\begin{displaymath}
    u^\alpha \left( \nabla_\alpha \left(\frac{\omega}{n} u_{\beta} \right) - 
    \nabla_\beta \left(\frac{\omega}{n} u_{\alpha} \right)  \right)=0.
\end{displaymath}

\noindent or:

\begin{equation}
  u^\alpha \left( \frac{\partial}{\partial
  x^{\alpha}}\left(\frac{\omega}{n} u_{\beta} \right) -
  \frac{\partial}{\partial x^{\beta}}\left(\frac{\omega}{n} u_{\alpha}
  \right)  \right)=0.
    \label{f-s}
\end{equation}

As shown by \citet{mendoza2021matter},
the matter Lagrangian of an ideal fluid 
is obtained from the following three conditions: (i) the definition
of the Hilbert energy-momentum tensor given by equation~\eqref{em-co},
(ii) the value~\eqref{energy-momentum-ideal} of the energy-momentum tensor for
an ideal fluid and (iii) the variations with respect to the metric tensor 
of the continuity
equation~\eqref{continuity}\footnote{

  Let us consider  a conservation equation of the form:

\begin{equation}
  \nabla_\alpha \left( A^\alpha \right)=0.
\end{equation}

\noindent Integrating this relation over an arbitrary 4-hypersurface
\( \Omega \) and using Gauss's theorem, it follows that:

\begin{equation}
 \int \mathrm{d}{\Omega} A^{\beta}_{; \beta} \sqrt{-g}  =
 \ointclockwise \mathrm{d}S_{\beta} A^{\beta} \sqrt{-g} = 0,
\end{equation}

\noindent over any  3-hypersurface with 3-volume element vector \( \mathrm{d}
S_{\beta} \) that bounds the arbitrary 4-hypersurface \( \Omega \) which
has a 4-volume element \( \mathrm{d} \Omega \).  So, performing the
variation of the fixed 3-volume integral it follows that:

\begin{displaymath}
    \delta \ointclockwise \mathrm{d}S_{\beta} A^{\beta} \sqrt{-g} =
    \ointclockwise \mathrm{d}S_{\beta} \delta \left( A^{\beta} \sqrt{-g}
    \right) = 0.
\end{displaymath}

Since this result is valid for any $3$-dimensional hypersurface which
bounds any given arbitrary $4$-dimensional surface $\Omega$, then
necessarily:

\begin{equation}
    \delta \left( A^{\beta} \sqrt{-g} \right) = 0.
\end{equation}

}:

\begin{equation}
  \delta \left( \sqrt{-g} n u^\alpha \right) = 0.
\label{var-continuity}
\end{equation}

\noindent to finally obtain\footnote{The sign on equation~\eqref{lme} also depends on the chosen signature
for the metric.  With a chosen signature (\(-,+,+,+\))
for the metric,  the matter Lagrangian would be:

\begin{displaymath}
  \lmatt = \pm e,
\end{displaymath}

\noindent and the plus-minus will also depend on the definition of the
matter action~\eqref{action}.}
:

\begin{equation}
    \lmatt=\mp e,
    \label{lme}
\end{equation}

\section{Potential flow non-coupled to the gravitational field}
\label{non-coupled-potential-flow}

  The first calculation that involves an ideal  non-coupled
potential flow  to the gravitational field was performed
by~\citet{taub1969stability}.  In Appendix~\ref{taub-current} we have
reproduced such calculations using modern variational techniques. 
Later on, it was~\citet{schutz70} who showed that the matter
Lagrangian of a general potential flow non-coupled to the gravitational
field is also given by this value.

  Concerning ideal potential flows non-coupled with the gravitational field
that reproduce the equations of hydrodynamics, they are all essentially
an extension of the non-relativistic pioneering work made
by~\citet{clebsh}, later on improved by the work of
\citet{khalatnikov52} with the general relativistic generalisations made
by~\citet{davydov49}, \citet{lin63}, \citet{taub01}, \citet{taub59},
and the most general case made by \citet{schutz70}.

  The value of the matter Lagrangian has been
obtained for a potential flow\footnote{It is important to remark that in the most general case, an
ideal fluid must be described as
the sum of a potential term (non-rotational term) plus a rotational one
according to De Rham's decomposition theorem~\citep[cf][]{nakahara}.} using the fact that the velocity potential
is independent of the metric tensor, with the end result that $\lmatt = \pm p$
\cite{taub1969stability, hawking-ellis, haghani2024first}.  

 As such, the value of the matter Lagrangian has been historically presented
as a degenerate value between the total energy density $e$ and the
pressure $p$, where both values yield the same energy-momentum tensor
\cite{de1990relativity, haghani2024first}.

  To understand why the matter Lagrangian can present a pressure value, let 
us look at the simplest potential flow
solution as described first by \citet{khalatnikov54}.  To do so, note
that due to the symmetries contained in the parenthesis on the left-hand
side of Euler's equation~\eqref{f-s}, imply that the term in
parenthesis is null when:

\begin{equation}
  u_\alpha = \frac{n}{\omega} \frac{ \partial \phi}{ \partial x^\alpha},
\label{vel-pot}
\end{equation}

\noindent for a scalar potential \( \phi \)\footnote{The formal way to
obtain solution~\eqref{vel-pot} is given in Appendix~\ref{vvel-pot}.}.
In other words, this
particular solution is valid only for an irrotational flow. With this
and using condition~\eqref{normalu} it follows that:

\begin{equation}
   \zeta ^{1/2} :=   \frac{\omega}{n}=\sqrt{g^{\alpha \beta}
     \frac{\partial \phi}{\partial x ^\alpha} \frac{\partial \phi}{\partial
     x ^\beta}},
\end{equation}

\noindent so that:

\begin{equation}
    u_\beta =  \frac{1}{\zeta^{1/2}} \frac{\partial \phi}{\partial x ^\beta}
    \label{khalatnikov}
\end{equation}

\noindent where:

\begin{equation}
    \zeta = \left(\frac{\omega}{n} \right)^2.
    \label{zeta}
\end{equation}

 If the matter Lagrangian is such that~\citep{avelino18}:

\begin{equation}
  \lmatt = \lmatt(\phi, \zeta),
\label{a01}
\end{equation}

\noindent where:

\begin{equation}
    \quad \zeta := | \boldsymbol{\nabla} \phi  |^2 = 
    \nabla^\alpha \phi \nabla_\alpha \phi,
\end{equation}

\noindent and if Khalatnikov's potential $\phi$ and its derivatives do not depend on the metric tensor $g^{\alpha \beta}$, from equation \eqref{a01} it
follows that \cite{avelino18}:

\begin{equation}
  T_{\alpha\beta} = \pm \frac{ \partial{ \lmatt } }{ \partial \zeta }
  \nabla_\alpha \phi \nabla_\beta \phi \mp \lmatt g_{\alpha \beta}.
\label{a02}
\end{equation}

  Direct comparison of equations~\eqref{a02}
and~\eqref{energy-momentum-ideal} imply that $\lmatt= \pm p$.

  Schutz's general proposal  for the representation of velocity
potentials is \cite{schutz70}:

\begin{equation}
    u_\nu = \frac{n}{\omega} (\partial_\nu \phi + \alpha \partial_\nu \beta + \theta \partial_\nu s),
\label{vel-schutz}
\end{equation}

\noindent where \( \alpha \), \( \beta \), \(\theta\), \( \phi \) and \(
s \) are velocity potentials (scalar fields)~\citep{de1990relativity}.
The choice~\eqref{vel-schutz} makes the parenthesis
in the left-hand side of~\eqref{f-s} null.  In other words, this
velocity is a 
solution of Euler's equation for the specific case of a general potential flow.  Note
that the first term on the right-hand side of
equation~\eqref{vel-schutz} is Khalatnikov's solution~\eqref{vel-pot}.
Since all
Schutz's potentials are additive and their individual variations are
null, their addition  lead to the same pressure
value for the matter Lagrangian in a very similar form as the one
presented above using Khalatnikov's potential.

  An alternative modern development to obtain the matter Lagrangian
consists on using the fact that the variations of Taub's
current~\cite{haghani2024first, de1990relativity}:

\begin{equation}
  V_\mu := (\omega/n) u_\mu,
\label{taubcurrent}
\end{equation}

\noindent where $u_\mu$
is given by equation \eqref{vel-schutz}, is null.  This is a
consequence of the general Lagrangian since all Schutz's potentials are
Lagrange multipliers.

In general, by the first law of thermodynamics~\eqref{flw}, the variation of the pressure with respect to the metric is given by:

\begin{equation}
    \delta p = n \delta \left( \frac{\omega}{n} \right).
    \label{deltapfl}
\end{equation}

In this case, if Schutz's six velocity potentials and their derivatives
are independent of the metric tensor, the variation of Taub's current
$V_\mu$ with respect to the metric tensor is zero,
and so:

\begin{equation}
    \delta \left( \frac{\omega}{n} \right) =  -\left( \frac{\omega}{n} \right) u^\mu \delta u_\mu.
\label{deltav0}
\end{equation}

From the normalisation condition~\eqref{normalu},
it follows that $\delta (g^{\mu \nu}u_\mu u_\nu)=0$ and, since the metric
tensor $g^{\mu \nu}$ is symmetric, we can write:

\begin{equation}
    u^\mu \delta u_\mu= - \frac12 u_\mu u_\nu \delta g^{\mu \nu}.
\label{deltau1}
\end{equation}

From equations \eqref{deltapfl}-\eqref{deltau1}, it follows that:

\begin{equation}
    \frac{\delta p}{\delta g^{\mu \nu}} = \frac12 (e+p) u_\mu u_\nu.
    \label{vp-harko}
\end{equation}

Equations \eqref{vp-harko}, \eqref{em-co} and
\eqref{energy-momentum-ideal} 
imply that $\lmatt = \pm p$ \cite{haghani2024first, de1990relativity}.

\section{Potential flow coupled to the gravitational field}
\label{coupled-potential-flow}

Now, if we assume that Khalatnikov's potential $\phi$ and its
derivatives depend on the metric tensor, then:

\begin{equation}
    \frac{\delta \lmatt}{\delta g^{\mu \nu}} = \frac{\partial \lmatt}{\partial \zeta} \frac{\delta \zeta}{\delta g^{\mu \nu}},
    \label{ac1}
\end{equation}

\noindent where by means of the first law of thermodynamics~\eqref{flw}:

\begin{equation}
    \frac{\delta \zeta}{\delta g^{\alpha \beta}} = 2 \frac{\omega}{n^2}\frac{\delta p}{\delta g^{\alpha \beta}},
    \label{ac2}
\end{equation}

\noindent and:

\begin{equation}
    \frac{\partial \lmatt}{\partial \zeta} = \frac12 \frac{n^2}{\omega} \frac{\partial \lmatt}{\partial p}.
    \label{dlz}
\end{equation}

With equations \eqref{ac1}-\eqref{dlz}, the Hilbert energy-momentum tensor \eqref{em-co} can be written as:

\begin{equation}
    T_{\alpha \beta} = \pm  2  {\pdv{\lmatt }{p}}  \frac{\delta p}{\delta g^{\alpha \beta}} \mp
   g_{\alpha\beta} \lmatt.
   \label{emtlp}
\end{equation}

\noindent Therefore, in this case, the value of the Hilbert
energy-momentum tensor depends on $\delta p / \delta g^{\alpha
\beta}$. Equation \eqref{emtlp} is consistent with the general case where
the chosen thermodynamic variable  is $p$ for an ideal fluid.
To obtain a full description
of the fluid  it is necessary to include an
equation of state~\cite{garcia1976introduccion} such as
relation~\eqref{eof} or~\eqref{eof1}.

 Under the assumption that Khalatnikov's potential $\phi$ and its 
derivatives depend on the metric tensor, it then follows
from~\eqref{khalatnikov}  that:

\begin{equation}
    \frac{\delta}{\delta g^{\alpha \beta}}\left( \frac{\omega}{n}
    u_{\mu}\right)=- \frac{\delta}{\delta g^{\alpha
    \beta}}\frac{\partial \phi}{\partial x^\mu}.
    \label{taubnonzero1}
\end{equation}

  From this relation it follows that 
the variation of Taub's current is in general different from zero and
so, expansion of that equation yields:

\begin{equation}
    \frac{\delta p}{\delta g^{\alpha \beta}} = \frac12 \omega u_\alpha u_\beta - nu^\mu \frac{\delta}{\delta g^{\alpha \beta}} \frac{\partial \phi}{\partial x^\mu }.
    \label{varp-acop}
\end{equation}

Since expression~\eqref{varp-acop} needs to be equivalent to an
equation of state of the form~\eqref{eof1}, or equivalently~\eqref{eof},
the following relation is necessary:

\begin{equation}
    u^\mu \frac{\delta}{\delta g^{\mu \nu} }\frac{\partial \phi}{\partial x^\mu}= \frac{1}{2} \frac{\omega}{n}u_\alpha u_\beta - \frac{1}{2} \frac{p \gamma}{n}(g_{\alpha \beta} - u_\alpha u_\beta),
\end{equation}

\noindent which can be rewritten as: 

\begin{equation}
    u^\mu \frac{\delta}{\delta g^{\mu \nu} } \frac{\partial
    \phi}{\partial x^\mu}= -u^\mu \frac{\delta}{\delta g^{\mu \nu} }
    V_\mu,
    \label{condition}
\end{equation}

\noindent with the aid of equation~\eqref{deltav0}.

Under condition~\eqref{condition}, equation~\eqref{varp-acop} implies that:

\begin{equation}
  \frac{\delta p}{ \delta g^{\alpha\beta}} = \frac{ 1 }{ 2 } p \gamma \left( g_{\alpha\beta} - u_\alpha
    u_\beta \right). 
\end{equation}

Consequently, condition~\eqref{condition} allows us to recover the
same variation of the pressure as the one obtained directly from the equation
of state~\eqref{deltap}. This ensures consistency between the result
obtained from the equation for an irrotational fluid and that obtained
from the equation of state of an ideal fluid.

Therefore, in this case $\lmatt=\mp e$ is the only solution, exactly
what was shown in Section~\ref{hydro}.

For the case of the six potentials defined by Schutz, the variation
of the addition most be equal to the right-hand side of 
equation~\eqref{condition}. In this general case it follows straightforward that
the matter Lagrangian is also given by~\eqref{lme}.

\section{General result}
\label{general-result}

  By the use of equation~\eqref{var-continuity}, the variation of the
particle number density is given by:

\begin{equation}
  \frac{\delta n}{ \delta g^{\alpha\beta}} = \frac{ 1 }{ 2 } n \left(
    g_{\alpha\beta} - u_\alpha u_\beta \right).
\label{deltarho}
\end{equation}

  Also, from the first law of thermodynamics~\eqref{fleos}, the
variations of the pressure are given by:

 \begin{equation}
        \delta p = \gamma \frac{p}{n}\delta n.
\label{chain-rule}  
\end{equation}  

Substituting equation \eqref{deltarho} into equation \eqref{chain-rule}, we find that:

\begin{equation}
  \frac{\delta p}{ \delta g^{\alpha\beta}} = \frac{ 1 }{ 2 } p \gamma \left( g_{\alpha\beta} - u_\alpha
    u_\beta \right). 
\label{deltap}
\end{equation}

Equation \eqref{deltap} is consistent with the variation of the equation of state of an ideal fluid \eqref{eof1} \footnote{

The direct variation of the pressure from the equation of
state~\eqref{eof1} is:

\begin{equation}
    \delta p = (\delta e - m \delta n)(\gamma - 1).
    \label{footdp}
\end{equation}

\noindent From equations~\eqref{flif} and~\eqref{deltarho} it follows
that:

\begin{equation}
    \delta e = \frac12 (e+p)(g_{\alpha \beta} - u_\alpha u_\beta) \delta g^{\alpha \beta}.
    \label{deltae}
\end{equation}

Substitution of  expressions~\eqref{eof1},~\eqref{deltarho}
and~\eqref{deltae} into equation~\eqref{footdp}, the same
result as that of equation \eqref{deltap} is obtained.  }.

Using the variation of the pressure given in equation~\eqref{deltap},
the Hilbert energy-momentum tensor in the representation of the pressure 
$p$ given by equation~\eqref{emtlp} can be written as:

\begin{equation}
 T_{\alpha\beta} = \pm p \gamma \left( g_{\alpha\beta} - u_\alpha
    u_\beta \right) \pdv{\lmatt }{p} \mp
   g_{\alpha\beta} \lmatt.
   \label{tl}
\end{equation}

Direct comparison between the energy momentum tensor of an ideal fluid
given by equations~\eqref{energy-momentum-ideal} and equation ~\eqref{tl} implies that:

\begin{equation}
\begin{split}
  \left( e + p \right) u_\alpha u_\beta - p g_{\alpha\beta} &= 
 \left(\pm p \gamma \pdv{\lmatt }{p}
    \mp \lmatt \right) g_{\alpha\beta} 
    \\
    &\mp p \gamma u_\alpha u_\beta  \pdv{\lmatt }{p},
    \end{split}
\label{diff-equation}  
\end{equation}

\noindent whose unique solution is relation~\eqref{lme}.

Therefore, in the general case of an ideal fluid where $\lmatt=\lmatt(p)$
the unique value of the matter Lagrangian is $\lmatt=\mp e$. Additionally,
in the case of a potential fluid given by $\lmatt=\lmatt(\zeta)$,
if $\zeta$ depends on $g^{\mu \nu}$ as shown above,
the unique solution $\lmatt=\mp e$ still remains.

\section{Discussion}
\label{discussion}

If we introduce the information of the equation of state~\eqref{eof1},
or equivalently~\eqref{eof}, into the equation of a 
irrotational flow~\eqref{khalatnikov} and under the  assumption  that
Khalatnikov's potential~$\phi$
and its derivatives do not depend on the metric tensor, then the following
relation for the variation of the pressure is obtained:

\begin{equation}
\frac{\delta p}{\delta g^{\alpha \beta}} = \frac12 p(g_{\alpha \beta}
-u_\alpha u_\beta) + \frac12 \omega \frac{\gamma -1}{\gamma} u_\alpha
u_\beta.
\label{varp_eos}
\end{equation}

 This expression implies that the matter Lagrangian is given by:

\begin{equation}
    \lmatt = \frac{p \omega}{\omega \left( \gamma - 1 \right) / \gamma  - p} +p,
\end{equation}

\noindent where a solution of the form $\lmatt=p$ implies
$p=-e$, i.e. $\lmatt=p=-e$. 
Additionally, we can do a direct comparison between
equations \eqref{deltap} and \eqref{varp_eos} to notice that both expressions are
equivalent if and only if $p=0$.

With all this,  we can see that if $\phi$ and its derivatives do not
depend on the metric tensor, to introduce the equation of state into the
equation of an irrotational flow implies $\lmatt = p = -e =0$. Therefore,
the energy density is given by $e=-p=0$ which by the equation of
state~\eqref{eof1} means that $n=0$ and so,
there are no fluid particles in the space.

  The calculations presented in this article imply that the value of the
matter Lagrangian of an ideal fluid is always given by its total
energy density as shown in equation~\eqref{lme}, which is fully consistent
with the first law of thermodynamics~\eqref{flif}, the null divergence of
the energy-momentum tensor~\eqref{conservation} and the following
unavoidable physical Principle (for the case of potential flows):

\begin{quote}
  \textbf{Principle.} \textit{An ideal fluid for which its
    4-velocity can be expressed as the
gradient of scalar potentials, must be coupled to the metric tensor. 
}
\end{quote}

\section*{Acknowledgements} 
This work was supported by PAPIIT DGAPA-UNAM grants IN110522 and
IN118325. SM and SS acknowledge support from Sechti (26344,1228643).

%%%%%%%%%%%%%%%%
% BIBLIOGRAPHY %
%%%%%%%%%%%%%%%%
\bibliographystyle{apsrev4-2}
\bibliography{mendoza-silva}

\appendix

\section{Taub's (1969) pressure calculation}
\label{taub-current}

 For the case of an ideal fluid, let us assume a constant gravitational
field~\citep{landau-fluids}, i.e. $\partial g_{\alpha \beta} / \partial
x^0 =0$, so that the fluid's 3-velocity is given by:

\begin{equation}
    v=\frac{\mathrm{d}l}{\mathrm{d}\tau}=\frac{c \mathrm{d}l }{\sqrt{g_{00}} \mathrm{d}x^0},
    \label{v1}
\end{equation}

\noindent and,

\begin{equation}
    u^k = \Gamma \frac{v^k}{c}, 
    \label{u1}
\end{equation}

\noindent where $\Gamma := 1/ \sqrt{1-v^2 / c^2}$. From~\eqref{v1} 
and \eqref{u1}, it follows that:

\begin{equation}
    u^0 = \Gamma \left( \frac{1}{\sqrt{g_{00}}} - \frac{g_{ok} v^k}{g_{00}c} \right). 
\end{equation}

To present the result originally derived by \citet{taub1969stability} in modern notation, we will work in the proper reference frame of the fluid, where $v^k = u^k=0 $ and $u^0 = 1/ \sqrt{g_{00}}$, in this case, equation \eqref{ec-t-c} is given by:

\begin{equation}
    u^0 \nabla_0 \left( \frac{\omega}{n} u_0 \right) =  \frac{\partial}{\partial x^0} \left( \frac{\omega}{n}  \right),
\end{equation}

\noindent and,

\begin{equation}
    \frac{\partial}{\partial x^k} \left( \frac{\omega}{n}  \right)=0.
    \label{l1}
\end{equation}

Since $u_0 =\sqrt{g_{00}}$ and assuming that all thermodynamical 
quantities are independent of $x^0$, equation \eqref{l1} takes the form:

\begin{equation}
    \nabla_0 \left(\frac{\omega}{n} \sqrt{g_{00}} \right)=0,
\end{equation}

\noindent which implies:

\begin{equation}
     \frac{\omega}{n} \sqrt{g_{00}} = \text{const.}
     \label{bernoulli}
\end{equation}

Relation~\eqref{bernoulli} is the relativistic version of Bernoulli's
equation in a comoving coordinate system.

Taking the the ratio of the variation of equation~\eqref{bernoulli} to itself
yields:

\begin{equation}
    \frac{\delta e + \delta p }{e+p} + \frac12 \frac{\delta
    g_{00}}{g_{00}} = \frac{\delta n}{n},
    \label{ratio}
\end{equation}

\noindent which can be rewritten as:

\begin{equation}
    \delta p = - \frac{1}{2} (e+p) \frac{\delta g_{00}}{g_{00}},
    \label{taub-dp}
\end{equation}

\noindent with the aid of the first law of thermodynamics~\eqref{flif}.

Substitution of expression \eqref{taub-dp} into Hilbert-Einstein field
equations: 

\begin{equation}
    G_{\mu \nu}\delta g_{\mu } = -\kappa T_{\mu \nu} \delta g_{\mu \nu}, 
\end{equation}

\noindent where \( \kappa = 8 \pi G \), gives the following full
Lagrangian \cite{taub1969stability} for general relativity:

\begin{equation}
    \mathcal{L}=R+2\kappa  p,
\end{equation}

\noindent which implies $\lmatt=p$.

\section{Velocity potential}
\label{vvel-pot}

The expression in parenthesis in equation~\eqref{f-s} is null if
the tensor:

\begin{displaymath}
   \Pi_{\alpha \beta}= \frac{\partial}{\partial x^{\alpha}}\left(\frac{\omega}{n} u_{\beta} \right),
\end{displaymath}

\noindent is symmetric and so, \( \epsilon^{\alpha\beta\gamma\lambda}
\Pi_{\alpha\beta} = 0\). As such, the integration of this tensor over an
arbitrary fixed  2D open manifold $M$, with area element \( \mathrm{d}
f_{\gamma \lambda} \), is null, i.e.:

\begin{displaymath}
    \int \frac12 \epsilon^{\alpha \beta \gamma \lambda} \mathrm{d} f_{\gamma \lambda} \Pi_{\alpha \beta}=0,
    \label{int1}
\end{displaymath}

\noindent or since the dual element of area is given by:

\begin{displaymath}
    \mathrm{d}f^{* \gamma \lambda}= \frac12 \epsilon^{\gamma \lambda
    \alpha \beta} \mathrm{d} f_{\alpha \beta},
\end{displaymath}

\noindent then by Stokes' theorem, it follows that:

\begin{displaymath}
    \int_{M}\mathrm{d}f^{* \alpha \beta} \Pi_{\alpha \beta} =\int_{M} \mathrm{d}f^{* \alpha \beta} \frac{\partial}{\partial x^{\alpha}}\left(\frac{\omega}{n} u_{\beta}\right)= \frac12 \ointclockwise \mathrm{d}x^{\beta} \frac{\omega u_{\beta}}{n} =0,
\end{displaymath}

\noindent where the closed line integral in the previous equation 
is integrated over a curve bounding \( M \).  In other words:

\begin{displaymath}
    \ointclockwise \mathrm{d}x^{\beta} \frac{\omega u_{\beta}}{n} =0,
\end{displaymath}

\noindent which necessarily implies that the integrand in the previous
equation is a full differential, i.e.:

\begin{displaymath}
    \mathrm{d}x^{\beta} \frac{\omega u_{\beta}}{n} = - \mathrm{d} \phi,
\end{displaymath}

\noindent so that:

\begin{displaymath}
     u_{\beta} = - \frac{n}{\omega} \frac{\partial \phi}{\partial x
     ^\beta},
\end{displaymath}

\noindent where \( \phi \) is Khalatnikov's potential.

\end{document}